\documentclass[12pt]{article}

\usepackage{fancyhdr}
\usepackage{amsmath}
\usepackage{amsfonts}
\usepackage{latexsym}
\usepackage{epsfig}
\usepackage{color}

\RequirePackage[text={165mm,222mm},centering,includefoot]{geometry}
\RequirePackage{layout}

\newcommand{\TR}{^{\rm T}}
\newcommand{\I}{^{\rm I}}
\newcommand{\ND}{^{({\rm n})}}
\newcommand{\MU}{^{(\mu)}}

\begin{document}

\newcounter{INDEX}
\setcounter{section}{0}
\setcounter{subsection}{0}
\setcounter{equation}{0}

\title{
The uniqueness of the invariant polarisation--tensor field {\color{black}for spin-1 particles in storage rings}} 

\author{D.~P.~Barber$^a$, A.~Kling$^b$ and M.~Vogt$^a$ \\
$^a$ \small{Deutsches~Elektronen--Synchrotron, DESY, ~22607 ~Hamburg,~Germany}
\\
$^b$ \small{ Institute of dual cooperative studies, Osnabrueck University of Applied Sciences,} \\
     \small{49809 Lingen, Germany    } }

\date{}

\maketitle


\begin{abstract} 
  We {\color {black} argue} that the invariant tensor field introduced in \cite{baritf2009} is unique
  under the condition that the invariant spin field is unique, and thereby complete that part of the 
  discussion in that paper.
\end{abstract}

\newpage


\newpage

\section{Introduction} 
\setcounter{equation}{0} 
In \cite{baritf2009}, where the invariant
tensor field (ITF) is introduced as a tool for describing equilibrium
spin-1 systems in storage rings, the matter of the uniqueness of the
ITF is mentioned and a plausible ansatz consistent with the 
definition of the ITF, the results of numerical
experiments, and with quantum mechanics is suggested. However, 
\cite{baritf2009} provides no rigorous
mathematical discussion of the topic and of course, we wish
to know whether more than one equilibrium spin-density-matrix field
can exist.  {\color {black} In this paper we close that gap by {\color {black} arguing}  that the ITF must be given
in terms of the invariant spin field, $\hat n$, by the ansatz
\begin{eqnarray}
T^{\rm I} = \pm {\sqrt{\frac{3}{2}}} \left \{ \hat n {\hat n}\TR -
  \frac{1}{3} I_{3 \times 3} \right \} \, ,
\label{eq:1.1}
\end{eqnarray}
of \cite{baritf2009} and is therefore unique 
up to a global sign under the condition that the invariant spin field
(ISF) is unique. }
Our approach provides insights into some necessary mathematical aspects of the ITF and it is
inspired by work on techniques for visualising the evolution of
tensors along (say) streamlines in fluids. These
techniques often exploit the fact that the properties of the relevant
tensors are completely encoded in their eigenvalues and eigenvectors
and that the tensors can be reconstructed from these latter. In our
case we study the evolution of the eigenvectors of real, $3
\times 3$, symmetric, Cartesian polarisation tensors along particle trajectories.
See \cite{visual} as an example of the copious literature on visualisation for tensors.

We begin with the definitions of the ITF and ISF and then recall some well
known properties of real symmetric matrices.  We are then in a position to
study the evolution of the eigenvectors along trajectories and arrive
at our proof.

It will be assumed that the reader is familiar with \cite{baritf2009}
and the context, and with the basic concepts of linear algebra.  Then,
apart from recalling the definitions of the ITF and the ISF, we shall
give no further introduction to the subject. The notation and ordering of coordinate axes will be the
same as in \cite{baritf2009}.

\setcounter{equation}{0}
\section{The definitions of the ITF and the ISF}
The ISF $\hat n (u; s)$ is a real 3-vector field with unit norm obeying the T-BMT 
equation along particle trajectories and it therefore evolves along a trajectory $u(s)$ as
\begin{eqnarray}
{\hat n}({M}(u; {\tilde s}, s); {\tilde s}) = R(u; {\tilde s}, s) {\hat n}(u; s) \; ,
\label{eq:0.1}
\end{eqnarray}
where ${M}(u; {\tilde s}, s)$ is the position in phase space at
$\tilde s \ge s$ after starting at $u$ and $s$, and $R(u; {\tilde s},
s)$ is the corresponding orthogonal transfer matrix representing the
solution to the T--BMT equation.

The ISF  also fulfills the periodicity condition
\begin{eqnarray}
{\hat n}(u; s + C) = {\hat n}(u; s) \, ,
\label{eq:0.2}
\end{eqnarray}
where $C$ is the circumference of the ring.

The ITF is a real, traceless, $3\times 3$, symmetric, Cartesian 
tensor field $T\I (u; s)$ evolving as
\begin{eqnarray}
T\I ( {M}(u; {\tilde s}, s); {\tilde s}) = R(u; {\tilde s}, s) \; T\I (u; s) R\TR(u; {\tilde s}, s) \;  ,
\label{eq:0.3} 
\end{eqnarray}
along a particle trajectory, and fulfilling the {\color {black} (same) } periodicity condition 
\begin{eqnarray}
T\I (u; s + C) = T\I (u; s) \, .
\label{eq:0.4} 
\end{eqnarray}
By definition
{\color {black}
\begin{eqnarray}
{\rm Tr}(T^{\I}) = 0 \, , 
\label{eq:0.5} 
\end{eqnarray}
and we normalise so that 
\begin{eqnarray}
{\mathfrak T}\I \equiv  \sqrt{{\rm Tr}( T^{\I} T^{\I} ) } = 1.
\label{eq:0.6} 
\end{eqnarray}
}
Away from orbital resonances and spin-orbit resonances, $\hat n (u; s)$
is unique up to a global sign \cite{beh2004}. In the following we {\color {black} argue}  that under the same conditions, 
the ITF is unique too and that it is given by (\ref{eq:1.1}).

\setcounter{equation}{0}
\section{The eigenspectra for real symmetric matrices}

We continue by recalling the eigenspectra of real, $j \times j$,
symmetric matrices. We do not insist that they are traceless {\color {black} at this stage}.

A real symmetric $j \times j$ matrix $A$ always has $j$ real eigenvalues $\Lambda$,
and $j$ real eigenvectors $E$ \cite{st1965}
so that
\begin{eqnarray}
A E_i = \Lambda_i E_i \, , \quad i = 1 \ldots j \, .
\nonumber 
\end{eqnarray}

In general the eigenvalues can be degenerate but even with degeneracy
the eigenvectors can be chosen to be mutually orthogonal.  Moreover,
it can be shown that these eigenvectors are complete in that they form
a basis in the $j$-dimensional vector space. Of course, the
eigenvectors can be scaled to have unit norms.  Then we can write
\begin{eqnarray}
A = U D U\TR \, ,
\label{eq:1.4} 
\end{eqnarray}
where $D$ is the $j \times j$  diagonal matrix of the eigenvalues 
and $U$ is an orthogonal $j \times j$ matrix whose columns  are
the eigenvectors {\color {black} $E_l$  $(l = 1 \ldots j)$ .} 
This relation can also be written as a spectral decomposition   
in terms of the  projectors $E_l \, E\TR_l$:
\begin{eqnarray}
A =  \sum_{l = 1}^{j} \Lambda_l  \, E_l \, E\TR_l  \, ,
\label{eq:1.5}  
\end{eqnarray}
and since the matrix $U$ is orthogonal we also have
\begin{eqnarray}
I =  \sum_{l = 1}^{j} E_l \, E\TR_l  \, ,
\label{eq:1.6}  
\end{eqnarray}
where $I$ is the $j \times j$ unit matrix.

For a $3 \times 3$ real, symmetric matrix $A$ with eigenvalues $\Lambda_i$ and
eigenvectors $E_i$, we have
\begin{eqnarray}
A E_i = \Lambda_i E_i \, , \quad i = 1,2,3 \, .
\nonumber 
\end{eqnarray}
where we again set the norms of the $E_i$ to unity so that they are
``unit vectors''.  
Then there are three cases:

\vspace{5mm}
\noindent
{\underline {\em Case 1: Three different eigenvalues}}

The three normalised eigenvectors $E$ are unique up to signs and form an orthonormal basis.
\vspace{5mm}

\noindent
{\underline{\em Case 2: Two  equal eigenvalues} }

We denote the common eigenvalue by $\mu$ and the two corresponding
eigenvectors by $E^{(\mu)}_1$ and $E^{(\mu)}_2$. We denote the
non-degenerate eigenvalue by $\Lambda \ND$ and its eigenvector by $E
\ND$. This is unique up to a sign. $E^{(\mu)}_1$ and $E^{(\mu)}_2$ are orthogonal to $E
\ND$ and can be chosen to be orthogonal to each other so that we again
have an orthonormal basis. However, this is not unique since any linear 
combination of $E^{(\mu)}_1$ and $E^{(\mu)}_2$ is still
an eigenvector to the eigenvalue $\mu$. In particular, the 
vectors obtained by rotating $E^{(\mu)}_1$ and $E^{(\mu)}_2$ together
around $E \ND$ by some angle $\theta$  are still
eigenvectors to the eigenvalue $\mu$:
\begin{eqnarray}
\left( \begin{array}{c} E^{\mu}_1 \\ E^{\mu}_2 \end{array}\right)_{\rm new}  =
\left( \begin{array}{cc} \ \ \  \cos \theta & \  \sin \theta\\ -\sin \theta & \,\, \cos \theta\end{array}\right) \left( \begin{array}{c} E^{\mu}_1 \\ E^{\mu}_2 \end{array}\right)_{\rm original} \; ,
\label{eq:1.7}
\end{eqnarray}
and we still have an orthonormal basis.
This is the case of most interest in this study.

For later use it is more convenient to cast (\ref{eq:1.7}) in complex form. Then we have 
\begin{eqnarray}
\left( E^{\mu}_1 + i E^{\mu}_2 \right )_{\rm new}  &=& e^{- i \theta}
\left( E^{\mu}_1 + i E^{\mu}_2 \right )_{\rm original} \nonumber \\
\left( E^{\mu}_1 - i E^{\mu}_2 \right )_{\rm new}  &=& e^{+ i \theta}
\left( E^{\mu}_1 - i E^{\mu}_2 \right )_{\rm original} \; .
\label{eq:1.8}
\end{eqnarray}

The normalised eigenvectors comprising an orthonormal basis are usually called principal axes. They provide coordinate systems
in which the tensor is diagonal.

\vspace{5mm}

\noindent
{\underline{\em Case 3: Three equal eigenvalues $\mu$} } 

{\color {black} In this case we can still arrange that the eigenvectors are mutually orthogonal 
so that the  matrix $U$ is orthogonal but otherwise arbitrary.
Now, with (\ref{eq:1.5}) and (\ref{eq:1.6}), we have $A = \mu I$.} This case is of no interest here since the 
unit matrix
cannot represent significant physics and $\mu$ must vanish anyway if $A$ is to be made traceless 
at some point.

\setcounter{equation}{0}
\section{The evolution of the eigenvectors of $T$}
We now examine the consequences of the previous section for a real, $3 \times 3$, symmetric, Cartesian 
tensor $T$ fulfilling the constraints (\ref{eq:0.3}) and  (\ref{eq:0.4}). 
We begin with Case 1.

\vspace{3mm}

For each $E_i(u; s) ~(i = 1,2,3)$ and from the orthogonality of $R$,
\begin{eqnarray}
 R(u; {\tilde s}, s) \; T (u; s) R\TR(u; {\tilde s}, s) R(u; {\tilde s}, s) E_i(u; s)
 = \Lambda_i  R(u; {\tilde s}, s) E_i(u; s)  \, .
\label{eq:4} 
\end{eqnarray}
so that 
\begin{eqnarray}
 T ({M}(u; {\tilde s}, s); {\tilde s}) R(u; {\tilde s}, s) E_i(u; s)
 = \Lambda_i  R(u; {\tilde s}, s) E_i(u; s) \; . 
\label{eq:5} 
\end{eqnarray}

Eigenvalues are invariant under similarity transformations such as that on the r.h.s. of (\ref{eq:0.3}).
Then, since in this case all of the eigenvalues $\Lambda$ are different, i.e.,
are non-degenerate, they provide unique labels for the eigenvectors and for each $i = 1,2,3$ we have 
\begin{eqnarray}
 R(u; {\tilde s}, s) E_i(u; s) = E_i ({M}(u; {\tilde s}, s); {\tilde s} ) \, ,
\label{eq:6} 
\end{eqnarray}
which is the eigenvector for the eigenvalue $\Lambda_i$ at the new
position ${M}(u; {\tilde s}, s)$ along the trajectory. 
It is then clear that the $E_i$ obey the T-BMT equation along trajectories.

Moreover, these $E_i$ are uniquely defined by the tensor $T$. They must therefore
exhibit the same periodicity as $T$ so that
\begin{eqnarray}
E_i (u; s + C) = E_i (u; s) \, , \quad i = 1,2,3 \, .
\label{eq:7}  
\end{eqnarray}
So, by definition, each $E_i$ is a vector $\hat n$. However, away from
orbital resonances and spin-orbit resonances the ISF, $\hat n(u; s)$, is
unique \cite{beh2004}.  We have therefore shown that the tensor field
$T$ subject to the constraints (\ref{eq:0.1}) and (\ref{eq:0.2})
cannot have three distinct eigenvalues away from orbital resonance and spin-orbit resonance.

We therefore consider Case 2. Here, as we have seen, the two
eigenvectors for the eigenvalue $\mu$ are orthogonal to $E\ND$ and can
be chosen to be orthogonal to each other, but they are not unique.
Then there are no unique relationships between the mutually-orthogonal eigenvectors
$E^{\mu}_1$ and $E^{\mu}_2$ chosen at some $u$ and $s$ and those chosen 
at $( {M}(u; {\tilde s}, s); \tilde s)$,
downstream along a trajectory. However, orthogonal transformations
preserve the angles between vectors so that we can still write
\begin{eqnarray}
 R(u; {\tilde s}, s) E^{(\mu)}_m(u; s) = \sum_{l = 1}^{2} a_{m l}  
E^{(\mu)}_l ({M}(u; {\tilde s}, s); {\tilde s}) \, ,
\label{eq:8} 
\end{eqnarray}
where the $2 \times 2$ matrix $a$ is orthogonal,

So the two eigenvectors are not constrained to satisfy the T-BMT equation. 
However, for the remaining
eigenvalue $\Lambda\ND $ we have 
\begin{eqnarray}
 R(u; {\tilde s}, s) E\ND (u; s) = E\ND ({M}(u; {\tilde s}, s); {\tilde s}) \, ,
\label{eq:9} 
\end{eqnarray}
and 
\begin{eqnarray}
E\ND (u; s + C) = E\ND (u; s) \, ,
\label{eq:10}  
\end{eqnarray}
so that away from orbital resonances and spin-orbit resonances $E\ND
(u; s) = \hat n (u; s)$.  Thus  the difficulty of Case 1 has been overcome
since the other two eigenvectors are not unique and need not 
obey the T-BMT equation. 
Moreover, they can, in fact, be chosen so that the set of vectors $E_l(u; s), E_2(u; s)$ and $E^{(n)}(u; s)$,
the principal axes,
forms a field of coordinate systems called the invariant frame field (IFF) \cite{beh2004,behII} although we do not need to do that here.
Within the coordinate system defined by the principal axes, $\hat n$ has the components $(0,1,0)$.

To arrive at the sought-after expression 
for $T$ we now return to  its spectral decomposition (\ref{eq:1.5}):
\begin{eqnarray}
T(u; s) &=&  \sum_{l = 1}^{3} \Lambda_l  \, E_l (u; s) \, E\TR_l (u; s) \nonumber \\
&=&
\Lambda\ND E\ND (u; s) \, E^{(n) {\rm T}} (u; s) + \mu \sum_{l = 1}^{2}
E\MU_l (u; s) \, E^{(\mu) {\rm T}}_l (u; s) \nonumber \\
&=&
\Lambda\ND \hat n (u; s) \, {\hat n}\TR (u; s) + \mu \sum_{l = 1}^{2}
E\MU_l (u; s) \, E^{(\mu) {\rm T}}_l (u; s) \, .
\label{eq:11}  
\end{eqnarray}

{\color {black}Of course, since $T$ is given, $\sum_{l = 1}^{2} E\MU_l (u; s) \,
E^{(\mu) {\rm T}}_l (u; s)$ must be invariant under the rotations (\ref{eq:1.7}) in the
plane orthogonal to $\hat n$. This can be confirmed by
writing}
\begin{eqnarray}
\sum_{l = 1}^{2} E\MU_l (u; s) \, E^{(\mu) {\rm T}}_l (u; s)  \qquad \qquad \qquad \qquad \qquad \qquad \qquad \qquad \qquad \qquad \qquad \qquad \qquad \qquad \nonumber \\
= \frac{1}{2} \lbrace
\left( E^{(\mu)}_1 + i E^{(\mu)}_2 \right ) \left( E^{(\mu)}_1 - i E^{(\mu)}_2 \right )\TR
+ \left( E^{(\mu)}_1 - i E^{(\mu)}_2 \right ) \left( E^{(\mu)}_1 + i E^{(\mu)}_2 \right )\TR \rbrace \; , 
\nonumber
\end{eqnarray}
and using (\ref{eq:1.8}).

Moreover, with (\ref{eq:1.6}) we have
\begin{eqnarray}
\hat n (u; s) \, {\hat n}\TR (u; s)   +  \sum_{l = 1}^{2}
E\MU_l (u; s) \, E^{(\mu) {\rm T}}_l (u; s) = I \, .
\label{eq:12}  
\end{eqnarray}

We now see that a real symmetric Cartesian tensor $T(u; s)$ fulfilling the requirements 
(\ref{eq:0.3}) and  (\ref{eq:0.4}) has the form 
\begin{eqnarray}
T(u; s) = (\Lambda\ND - \mu) \, \hat n (u; s) \, {\hat n}\TR (u; s)   +  \mu  \, I  \, .
\label{eq:13}  
\end{eqnarray}
In general a real, $3 \times 3$, symmetric tensor has six independent
parameters. These can be taken to be the three eigenvalues and the
three parameters defining the rotation embodied in the orthogonal $3
\times 3$ matrix $U$ in (\ref{eq:1.4}). However, the imposition of the
constraints (\ref{eq:0.3}) and (\ref{eq:0.4}) and some resulting necessary
degeneracy has reduced the number of parameters to four, namely
$\Lambda\ND, ~ \mu$, and two direction cosines of ${\hat n}$.

Next, by requiring that $T$ be traceless we obtain $\Lambda\ND = - 2 \mu$ so that
\begin{eqnarray}
T(u; s) = - 3 \mu \, \left ( \hat n (u; s) \, {\hat n}\TR (u; s)   -\frac{1}{3} \, I \right ) \, .
\label{eq:14}  
\end{eqnarray}
Then,  by requiring that ${\rm Tr}(T^2)$ = 1 we have $\mu = \pm 1/{\sqrt 6}$ so that we finally 
obtain  
\begin{eqnarray}
T^{\rm I}  = \pm   {\sqrt{\frac{3}{2}}} \left \{ \hat n {\hat n}\TR  - \frac{1}{3} I \right \} \, .
\label{eq:15}  
\end{eqnarray}
This has just two free parameters, namely two direction cosines of ${\hat n}$. 

It is simple to confirm with (\ref{eq:15}) 
that $\hat n$ is an eigenvector of $T^{\rm I}$ with the eigenvalue $\pm 2/{\sqrt{6}}$ and that the other two 
eigenvectors, orthogonal to $\hat n$, have the eigenvalue $\mp 1/{\sqrt{6}}$. 
The evolutions of rank-2, $3\times3$ tensors are often visualised with the aid of    
quadric surfaces whose major and minor axes are the principal axes mentioned earlier \cite{visual}.
In our case, ${\rm Tr} (T\I)$ vanishes so that the underlying quadratic form is not positive definite. In fact this   
quadric surface is a hyperboloid of two sheets invariant under rotation around $\hat n$. 
Since the eigenvalues do not vary along a particle trajectory,
the hyperboloid has a constant shape although its orientation changes as the direction of $\hat n$ changes.
The hyperboloid for a non-invariant $T_{\rm loc}$ changes shape as well as orientation along a particle trajectory
as its eigenvalues change.

With (\ref{eq:15}) we have arrived at our destination with the proof that the ITF,
namely a normalised, $3\times3$, real, symmetric, traceless, Cartesian
tensor fulfilling the requirements (\ref{eq:0.3}) and (\ref{eq:0.4}),
is unique up to a global sign if the ISF is unique, and that it then takes
the form (\ref{eq:15}).

The same conclusion is reached in a  broader context and in a more powerful manner in \cite[Section 8]{behII} 
whereby invariant fields are associated with symmetry groups
{\footnote {This paper was prepared some years before \cite{behII} but not distributed then.}}.
That work also addresses the case, on orbital resonance, when the ISF defined there
might not exist whereas an ITF can exist.
Our discussion on the number of distinct eigenvalues compliments the
discussion in \cite[Section 8]{behII}. In particular, we have also
found that spin-orbit resonance implies that the ITF has three distinct
eigenvalues. 

\section{Summary}
{\color {black}
We have augmented the work in \cite{baritf2009} to {\color {black} argue} that since the
ISF is unique away from spin-orbit resonances, the ITF is unique too,
up to a sign.  In contrast to the construction of the ITF in
\cite{baritf2009} where an appeal to semi-classical quantum mechanics
and numerical experiments was made, we have been able to rely on
purely mathematical arguments. The mathematical techniques used in  \cite{behII} set our result in a broader context.

\section*{Acknowledgments}
We thank J.A. Ellison and  K. Heinemann for fruitful collaboration on topics surrounding the ISF and the ITF.

\addcontentsline{toc}{section}{References}
\small \small

\end{document}